\documentclass{article}
\usepackage{amsmath}
\usepackage{amssymb}
\usepackage{enumitem}
\usepackage{csquotes}
\usepackage{array,arydshln} 
\usepackage{graphicx}
\usepackage{subfig}
\usepackage{url}
\usepackage{tikz}
	\newcommand{\adjustTikzSize}[0]{\Large}
	\newcommand{\tikzScale}[0]{0.65}
\usepackage{ntheorem}
	\theoremstyle{plain}
	\theorembodyfont{\upshape}
	\theoremsymbol{\ensuremath{\ast}}
	\theoremseparator{}

\newcommand{\Nat}{\mathbb{N}}
\newcommand{\R}{\mathbb{R}}
\date{}
\author{
	Frank Rosner\\
	Global Data and Analytics\\
	Allianz SE\\
	Germany
	\and 
	Alexander Hinneburg \\
	Computer Science\\
	Martin-Luther-University\\ Halle-Wittenberg\\ 
	Germany \\
}
    
\title{Translating Bayesian Networks into Entity Relationship Models\\ \large (Extended Version)\footnote{This document is an extended version of a paper published in the Proceedings of the 35th International Conference on Conceptual Modeling, ER 2016.
In addition to a more detailed discussion of the translation method presented in the conference version, this extended version provides a description of a case study that applies the method as well as first ideas of a conceptual framework for developing big data analytics applications.
}
}
\begin{document}
\maketitle
\begin{abstract}
Big data analytics applications drive the convergence of data management and machine learning.
But there is no conceptual language available that is spoken in both worlds.
The main contribution of the paper is a method to translate Bayesian networks, a main conceptual language for probabilistic graphical models, into usable entity relationship models.
The transformed representation of a Bayesian network leaves out mathematical details about probabilistic relationships but unfolds all information relevant for data management tasks.
As a real world example, we present the TopicExplorer system that uses Bayesian topic models as a core component in an interactive, database-supported web application.
Last, we sketch a conceptual framework that eases machine learning specific development tasks while building big data analytics applications.
\end{abstract}

\section{Introduction}

The implementation of a big data analytics application requires to join data management software with machine learning tools.
However, the fields of data management and machine learning developed quite different models and notations. The former frequently uses entity-relationship models (ERM) while the latter uses probabilistic graphical models to communicate key concepts.
In this paper, we pick Bayesian networks (BN) as a widely used graphical notation for machine learning models.
Note that the presented ideas can be transferred to other common graphical notions like undirected Markov models or factor graphs as well.

The notations of ERMs and BNs are designed to serve the needs of the respective fields. They are stressing information relevant in the particular domain while visually suppressing less important details.
E.g. an ERM highlights the existence and cardinalities of relationships between distinguishable entities.
On the other hand, a BN represents a joint probability distribution as a factorization of several hierarchically linked conditional probabilities.
Even while both kinds of graphical notations show many details of the data, information explicit on one side remains implicit on the other one and vice versa --- there is no natural understanding of the two worlds.
However, a common conceptual description of the contribution from both worlds is crucial for the successes of big data analytics projects.

The data management part of a big data analytics project typically represents more details of the data than the machine learning part.
Therefore, it is reasonable to translate the machine learning part to the conceptual language of data management.
Attempts into this direction include machine learning libraries with APIs in one or multiple programming languages \cite{sparks2013mli,hall2009weka,scikit-learn}, and new declarative languages or extension of existings ones \cite{Armbrust:2015:SSR:2723372.2742797,kumar2013hazy,domingos2007markov,akdere2011case,hellerstein2012madlib}.
However, none of these approaches solves the problem of integrating the information about machine learning that is relevant for data management into the conceptual view of this side.
The advantages of a formal conceptual view of machine learning models integrated into the conceptual view of the data management side would be (i) no black box behind an abstract API in the data management model and (ii) developers from the data management side understand the basic in- and outputs of the machine learning part.

We propose a rule-based method to translate a graphical BN model in plate notation into an entity relationship model.
Such ERM can be easily integrated into the overall ERM of the whole application.
As an example, we look at topic modeling of documents \cite{Blei:2012:PTM:2133806.2133826}.
Latent Dirichlet allocation (LDA) \cite{blei2003latent} -- shown in Figure \ref{fig:topic_platemodel_original} -- is one of the most popular topic models.
Although not limited to this application, it is often used to find a hidden structure in text documents, called \emph{topics}.
Topics are modeled as probability distributions over a vocabulary.
They are often presented as word lists each ordered by descending probabilities.

\begin{figure}[t]
\centering
\scalebox{0.5}{\adjustTikzSize \input{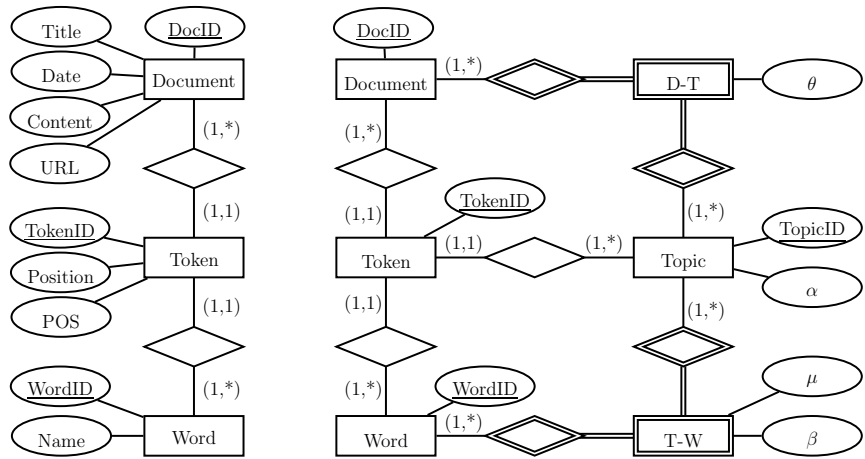}}
\caption{ERM of given Data (left) and translated ERM for LDA (right).}
\label{fig:ex1_er}
\end{figure}

The given data in this example include documents consisting of word tokens.
Each token corresponds to a particular occurrence of a word from the vocabulary.
Documents, tokens and words have additional information attached like title, publication date or part-of-speech (POS) tags.
The entities and relationships about the given data are shown in Figure \ref{fig:ex1_er} (left).
The BN describing the LDA topic model shown in Figure \ref{fig:topic_platemodel_original} represents the given word tokens $\vec d_{nm}$ as shaded circles, which indicates random variables with given values.
The hidden random variables represented by empty circles are the token-topic assignments $\vec z_{nm}$, document specific topic proportions $\vec\theta_n$ and topic-vocabulary distributions $\vec\mu_k$.
For those variables, either expectations or probable value assignments are computed during machine learning inference.
The black dots represent fixed hyper-parameters that determine the prior distributions for the hidden variables.
The descriptions of the boxes (plates) correspond to the entities \verb|Document| and \verb|Token|.
However, some entities like \verb|Word| do not appear in the BN as they are implicit in the definition of vectors.
Further, the plate named \verb|Topics| introduces a new entity.

The result of the whole translation is shown in Figure \ref{fig:ex1_er} (right).
The ERM of the BN represents all relevant entities with respective primary keys.
Further, the relationships between topics on one side and documents, tokens and vocabulary words on the other side are shown with their respective cardinalities.
The ERMs about the given data and the translated one from BN can be combined into a single one by merging the matching entities.
We believe that such an overall ERM helps to improve the efficiency of the development process as now the developers on the data management side can see what data is needed and contributed from the machine learning part with respect to relational aspects.

Our main contribution is a rule-based translation from BN in plate notation to ERMs, described in Section \ref{sec:pm2erm}.
Our method provides semantic guidelines for building a conceptual representation of a BN that addresses the needs of the data management side of a big data analytics project.
However, a BN is not a unique way to describe a probabilistic model, i.e. the same probabilistic model can be described by multiple BNs that differ in complexity.
Therefore, our proposed translation constructs an intermediate atomic plate model (APM) in several steps. It gradually uncovers implicit information not represented in the original BN.
Further, the subsequent translation from an APM to an ERM can include different probabilistic relationships between the generated entities.

We demonstrate the method in the real world example of the TopicExplorer in Section \ref{sec:casestudy}.
Based on this, we describe our vision of a conceptual framework that uses pre-translated BNs as a library of ERM snippets in Section \ref{sec:framework}.
Such library could be used by data management developers to conceptually include machine learning methods into analytics applications.
We believe that software development of big data analytics applications could benefit from machine learning implementations that are attached to the pre-translated BNs.
Last, we discuss related work in Section \ref{sec:relwork} and conclude the paper in Section \ref{sec:conclusion}.

\section{Translation of Bayesian Network in Plate notation}
\label{sec:pm2erm}
BNs mainly describe data with random variables at the level of data items, e.g. word tokens.
The BN for a topic model in Figure \ref{fig:topic_platemodel_original} has an observed random variable $\vec d_{nm}$ for each of the $M_n$ tokens of the $n$th document.
As plotting all data items one by one is not possible, the visual notation of plates is used.
A plate groups a set of random variables sharing an index set.
Due to this, plates convey information about entities and relationships.
However, some random variables are implicitly denoted in BNs: (i) multidimensional vector notation of random variables and (ii) functions that implicitly describe data transformation coupled with joins.
Therefore, we propose a stepwise approach to transform a given BN in plate notation into a well-formed ERM. This is done in three steps:
\begin{enumerate}
	\item Make implicit relational information explicit: The resulting model is called an atomic plate model (APM).
	\item\label{itm:apm2erm} Convert the APM into an ERM based on graphical rules.
	\item Reduce the ERM to avoid translation artifacts.
\end{enumerate}

We use the standard ERM notation \cite{chen1976entity} with min-max cardinalities \cite[p.~82]{elmasri2007database}.
We call an ERM \emph{well-formed} iff it (1) is syntactically correct, (2) is explicit and (3) does not have redundant constructs.
Explicitness means that all real world constructs which have a corresponding construct in the ERM notation are modeled using those.
Thus, a well-formed ERM does not contain explicit foreign key attributes but uses relationships instead.
Having no redundant constructs means that no entity or relationship is duplicate, i.e.~semantically expressing the same thing.
However, those duplicates may appear as intermediate results of the translation procedure.
The following subsections offer detailed explanations of all steps to translate a BN to an ERM.

\subsection{Construction of Atomic Plate Models}
Plate models (BN in plate notation) may contain multidimensional random variables (e.g. vectors or matrices), hidden deterministic data transformations, and relationships.
A plate model is converted to an APM by explicitly including those hidden transformations and relationships and any variables associated with it, as well as splitting multidimensional variables into their components.

\begin{figure}[t]
	\begin{minipage}[t]{0.49\linewidth}
		\begin{center}
	    \subfloat[Vector]{\label{fig:pm2apm_a}
           \scalebox{\tikzScale}{\adjustTikzSize 
\ifx\du\undefined
  \newlength{\du}
\fi
\setlength{\du}{15\unitlength}
\begin{tikzpicture}
\pgftransformxscale{0.809745}
\pgftransformyscale{-0.809745}
\definecolor{dialinecolor}{rgb}{0.000000, 0.000000, 0.000000}
\pgfsetstrokecolor{dialinecolor}
\definecolor{dialinecolor}{rgb}{1.000000, 1.000000, 1.000000}
\pgfsetfillcolor{dialinecolor}
\pgfsetlinewidth{0.100000\du}
\pgfsetdash{}{0pt}
\pgfsetdash{}{0pt}
\pgfsetmiterjoin
\definecolor{dialinecolor}{rgb}{0.000000, 0.000000, 0.000000}
\pgfsetstrokecolor{dialinecolor}
\draw (14.400000\du,0.800000\du)--(14.400000\du,5.800000\du)--(19.400000\du,5.800000\du)--(19.400000\du,0.800000\du)--cycle;
\definecolor{dialinecolor}{rgb}{0.000000, 0.000000, 0.000000}
\pgfsetstrokecolor{dialinecolor}
\node[anchor=east] at (19.594600\du,5.281300\du){$n \in N$};
\definecolor{dialinecolor}{rgb}{0.000000, 0.000000, 0.000000}
\pgfsetstrokecolor{dialinecolor}
\node[anchor=west] at (16.251400\du,2.300000\du){$x_n$};
\definecolor{dialinecolor}{rgb}{1.000000, 1.000000, 1.000000}
\pgfsetfillcolor{dialinecolor}
\pgfpathellipse{\pgfpoint{5.750000\du}{3.300000\du}}{\pgfpoint{0.500000\du}{0\du}}{\pgfpoint{0\du}{0.500000\du}}
\pgfusepath{fill}
\pgfsetlinewidth{0.100000\du}
\pgfsetdash{}{0pt}
\pgfsetdash{}{0pt}
\definecolor{dialinecolor}{rgb}{0.000000, 0.000000, 0.000000}
\pgfsetstrokecolor{dialinecolor}
\pgfpathellipse{\pgfpoint{5.750000\du}{3.300000\du}}{\pgfpoint{0.500000\du}{0\du}}{\pgfpoint{0\du}{0.500000\du}}
\pgfusepath{stroke}
\definecolor{dialinecolor}{rgb}{0.000000, 0.000000, 0.000000}
\pgfsetstrokecolor{dialinecolor}
\node[anchor=west] at (5.131100\du,2.210800\du){$\vec x$};
\definecolor{dialinecolor}{rgb}{1.000000, 1.000000, 1.000000}
\pgfsetfillcolor{dialinecolor}
\pgfpathellipse{\pgfpoint{1.250000\du}{3.300000\du}}{\pgfpoint{0.500000\du}{0\du}}{\pgfpoint{0\du}{0.500000\du}}
\pgfusepath{fill}
\pgfsetlinewidth{0.100000\du}
\pgfsetdash{}{0pt}
\pgfsetdash{}{0pt}
\definecolor{dialinecolor}{rgb}{0.000000, 0.000000, 0.000000}
\pgfsetstrokecolor{dialinecolor}
\pgfpathellipse{\pgfpoint{1.250000\du}{3.300000\du}}{\pgfpoint{0.500000\du}{0\du}}{\pgfpoint{0\du}{0.500000\du}}
\pgfusepath{stroke}
\definecolor{dialinecolor}{rgb}{0.000000, 0.000000, 0.000000}
\pgfsetstrokecolor{dialinecolor}
\node[anchor=west] at (0.631100\du,2.300000\du){$\alpha$};
\pgfsetlinewidth{0.100000\du}
\pgfsetdash{}{0pt}
\pgfsetdash{}{0pt}
\pgfsetbuttcap
{
\definecolor{dialinecolor}{rgb}{0.000000, 0.000000, 0.000000}
\pgfsetfillcolor{dialinecolor}
\pgfsetarrowsend{to}
\definecolor{dialinecolor}{rgb}{0.000000, 0.000000, 0.000000}
\pgfsetstrokecolor{dialinecolor}
\draw (1.750000\du,3.300000\du)--(5.250000\du,3.300000\du);
}
\definecolor{dialinecolor}{rgb}{1.000000, 1.000000, 1.000000}
\pgfsetfillcolor{dialinecolor}
\pgfpathellipse{\pgfpoint{16.900000\du}{3.300000\du}}{\pgfpoint{0.500000\du}{0\du}}{\pgfpoint{0\du}{0.500000\du}}
\pgfusepath{fill}
\pgfsetlinewidth{0.100000\du}
\pgfsetdash{}{0pt}
\pgfsetdash{}{0pt}
\definecolor{dialinecolor}{rgb}{0.000000, 0.000000, 0.000000}
\pgfsetstrokecolor{dialinecolor}
\pgfpathellipse{\pgfpoint{16.900000\du}{3.300000\du}}{\pgfpoint{0.500000\du}{0\du}}{\pgfpoint{0\du}{0.500000\du}}
\pgfusepath{stroke}
\definecolor{dialinecolor}{rgb}{1.000000, 1.000000, 1.000000}
\pgfsetfillcolor{dialinecolor}
\pgfpathellipse{\pgfpoint{12.400000\du}{3.300000\du}}{\pgfpoint{0.500000\du}{0\du}}{\pgfpoint{0\du}{0.500000\du}}
\pgfusepath{fill}
\pgfsetlinewidth{0.100000\du}
\pgfsetdash{}{0pt}
\pgfsetdash{}{0pt}
\definecolor{dialinecolor}{rgb}{0.000000, 0.000000, 0.000000}
\pgfsetstrokecolor{dialinecolor}
\pgfpathellipse{\pgfpoint{12.400000\du}{3.300000\du}}{\pgfpoint{0.500000\du}{0\du}}{\pgfpoint{0\du}{0.500000\du}}
\pgfusepath{stroke}
\definecolor{dialinecolor}{rgb}{0.000000, 0.000000, 0.000000}
\pgfsetstrokecolor{dialinecolor}
\node[anchor=west] at (11.781100\du,2.300000\du){$\alpha$};
\pgfsetlinewidth{0.100000\du}
\pgfsetdash{}{0pt}
\pgfsetdash{}{0pt}
\pgfsetbuttcap
\pgfsetmiterjoin
\pgfsetlinewidth{0.100000\du}
\pgfsetbuttcap
\pgfsetmiterjoin
\pgfsetdash{}{0pt}
\definecolor{dialinecolor}{rgb}{0.800000, 0.800000, 0.800000}
\pgfsetfillcolor{dialinecolor}
\fill (7.300000\du,3.050000\du)--(9.050000\du,3.050000\du)--(9.050000\du,2.800000\du)--(10.800000\du,3.300000\du)--(9.050000\du,3.800000\du)--(9.050000\du,3.550000\du)--(7.300000\du,3.550000\du)--cycle;
\definecolor{dialinecolor}{rgb}{0.800000, 0.800000, 0.800000}
\pgfsetstrokecolor{dialinecolor}
\draw (7.300000\du,3.050000\du)--(9.050000\du,3.050000\du)--(9.050000\du,2.800000\du)--(10.800000\du,3.300000\du)--(9.050000\du,3.800000\du)--(9.050000\du,3.550000\du)--(7.300000\du,3.550000\du)--cycle;
\pgfsetbuttcap
\pgfsetmiterjoin
\pgfsetdash{}{0pt}
\definecolor{dialinecolor}{rgb}{0.800000, 0.800000, 0.800000}
\pgfsetstrokecolor{dialinecolor}
\draw (7.300000\du,3.050000\du)--(9.050000\du,3.050000\du)--(9.050000\du,2.800000\du)--(10.800000\du,3.300000\du)--(9.050000\du,3.800000\du)--(9.050000\du,3.550000\du)--(7.300000\du,3.550000\du)--cycle;
\end{tikzpicture}}
	    }
		\end{center}
	\end{minipage}
	\hspace{0.0cm}
	\begin{minipage}[t]{0.49\linewidth}
		\begin{center}
	    \subfloat[Matrix]{\label{fig:pm2apm_b}
           \scalebox{\tikzScale}{\adjustTikzSize 
\ifx\du\undefined
  \newlength{\du}
\fi
\setlength{\du}{15\unitlength}
\begin{tikzpicture}
\pgftransformxscale{0.745638}
\pgftransformyscale{-0.745638}
\definecolor{dialinecolor}{rgb}{0.000000, 0.000000, 0.000000}
\pgfsetstrokecolor{dialinecolor}
\definecolor{dialinecolor}{rgb}{1.000000, 1.000000, 1.000000}
\pgfsetfillcolor{dialinecolor}
\pgfsetlinewidth{0.100000\du}
\pgfsetdash{}{0pt}
\pgfsetdash{}{0pt}
\pgfsetmiterjoin
\definecolor{dialinecolor}{rgb}{0.000000, 0.000000, 0.000000}
\pgfsetstrokecolor{dialinecolor}
\draw (36.950000\du,1.850000\du)--(36.950000\du,7.350000\du)--(42.450000\du,7.350000\du)--(42.450000\du,1.850000\du)--cycle;
\pgfsetlinewidth{0.100000\du}
\pgfsetdash{}{0pt}
\pgfsetdash{}{0pt}
\pgfsetmiterjoin
\definecolor{dialinecolor}{rgb}{0.000000, 0.000000, 0.000000}
\pgfsetstrokecolor{dialinecolor}
\draw (35.950000\du,0.850000\du)--(35.950000\du,6.350000\du)--(41.450000\du,6.350000\du)--(41.450000\du,0.850000\du)--cycle;
\definecolor{dialinecolor}{rgb}{0.000000, 0.000000, 0.000000}
\pgfsetstrokecolor{dialinecolor}
\node[anchor=east] at (41.644600\du,5.831300\du){$n \in N$};
\definecolor{dialinecolor}{rgb}{0.000000, 0.000000, 0.000000}
\pgfsetstrokecolor{dialinecolor}
\node[anchor=west] at (38.301400\du,2.850000\du){$x_{nm}$};
\definecolor{dialinecolor}{rgb}{1.000000, 1.000000, 1.000000}
\pgfsetfillcolor{dialinecolor}
\pgfpathellipse{\pgfpoint{27.250000\du}{3.850000\du}}{\pgfpoint{0.500000\du}{0\du}}{\pgfpoint{0\du}{0.500000\du}}
\pgfusepath{fill}
\pgfsetlinewidth{0.100000\du}
\pgfsetdash{}{0pt}
\pgfsetdash{}{0pt}
\definecolor{dialinecolor}{rgb}{0.000000, 0.000000, 0.000000}
\pgfsetstrokecolor{dialinecolor}
\pgfpathellipse{\pgfpoint{27.250000\du}{3.850000\du}}{\pgfpoint{0.500000\du}{0\du}}{\pgfpoint{0\du}{0.500000\du}}
\pgfusepath{stroke}
\definecolor{dialinecolor}{rgb}{0.000000, 0.000000, 0.000000}
\pgfsetstrokecolor{dialinecolor}
\node[anchor=west] at (26.631100\du,2.760800\du){$\vec X$};
\definecolor{dialinecolor}{rgb}{1.000000, 1.000000, 1.000000}
\pgfsetfillcolor{dialinecolor}
\pgfpathellipse{\pgfpoint{22.750000\du}{3.850000\du}}{\pgfpoint{0.500000\du}{0\du}}{\pgfpoint{0\du}{0.500000\du}}
\pgfusepath{fill}
\pgfsetlinewidth{0.100000\du}
\pgfsetdash{}{0pt}
\pgfsetdash{}{0pt}
\definecolor{dialinecolor}{rgb}{0.000000, 0.000000, 0.000000}
\pgfsetstrokecolor{dialinecolor}
\pgfpathellipse{\pgfpoint{22.750000\du}{3.850000\du}}{\pgfpoint{0.500000\du}{0\du}}{\pgfpoint{0\du}{0.500000\du}}
\pgfusepath{stroke}
\definecolor{dialinecolor}{rgb}{0.000000, 0.000000, 0.000000}
\pgfsetstrokecolor{dialinecolor}
\node[anchor=west] at (22.131100\du,2.850000\du){$\alpha$};
\pgfsetlinewidth{0.100000\du}
\pgfsetdash{}{0pt}
\pgfsetdash{}{0pt}
\pgfsetbuttcap
{
\definecolor{dialinecolor}{rgb}{0.000000, 0.000000, 0.000000}
\pgfsetfillcolor{dialinecolor}
\pgfsetarrowsend{to}
\definecolor{dialinecolor}{rgb}{0.000000, 0.000000, 0.000000}
\pgfsetstrokecolor{dialinecolor}
\draw (23.250000\du,3.850000\du)--(26.750000\du,3.850000\du);
}
\definecolor{dialinecolor}{rgb}{1.000000, 1.000000, 1.000000}
\pgfsetfillcolor{dialinecolor}
\pgfpathellipse{\pgfpoint{38.950000\du}{3.850000\du}}{\pgfpoint{0.500000\du}{0\du}}{\pgfpoint{0\du}{0.500000\du}}
\pgfusepath{fill}
\pgfsetlinewidth{0.100000\du}
\pgfsetdash{}{0pt}
\pgfsetdash{}{0pt}
\definecolor{dialinecolor}{rgb}{0.000000, 0.000000, 0.000000}
\pgfsetstrokecolor{dialinecolor}
\pgfpathellipse{\pgfpoint{38.950000\du}{3.850000\du}}{\pgfpoint{0.500000\du}{0\du}}{\pgfpoint{0\du}{0.500000\du}}
\pgfusepath{stroke}
\definecolor{dialinecolor}{rgb}{1.000000, 1.000000, 1.000000}
\pgfsetfillcolor{dialinecolor}
\pgfpathellipse{\pgfpoint{34.450000\du}{3.850000\du}}{\pgfpoint{0.500000\du}{0\du}}{\pgfpoint{0\du}{0.500000\du}}
\pgfusepath{fill}
\pgfsetlinewidth{0.100000\du}
\pgfsetdash{}{0pt}
\pgfsetdash{}{0pt}
\definecolor{dialinecolor}{rgb}{0.000000, 0.000000, 0.000000}
\pgfsetstrokecolor{dialinecolor}
\pgfpathellipse{\pgfpoint{34.450000\du}{3.850000\du}}{\pgfpoint{0.500000\du}{0\du}}{\pgfpoint{0\du}{0.500000\du}}
\pgfusepath{stroke}
\definecolor{dialinecolor}{rgb}{0.000000, 0.000000, 0.000000}
\pgfsetstrokecolor{dialinecolor}
\node[anchor=west] at (33.831100\du,2.850000\du){$\alpha$};
\pgfsetlinewidth{0.100000\du}
\pgfsetdash{}{0pt}
\pgfsetdash{}{0pt}
\pgfsetbuttcap
\pgfsetmiterjoin
\pgfsetlinewidth{0.100000\du}
\pgfsetbuttcap
\pgfsetmiterjoin
\pgfsetdash{}{0pt}
\definecolor{dialinecolor}{rgb}{0.800000, 0.800000, 0.800000}
\pgfsetfillcolor{dialinecolor}
\fill (28.700000\du,3.600000\du)--(30.450000\du,3.600000\du)--(30.450000\du,3.350000\du)--(32.200000\du,3.850000\du)--(30.450000\du,4.350000\du)--(30.450000\du,4.100000\du)--(28.700000\du,4.100000\du)--cycle;
\definecolor{dialinecolor}{rgb}{0.800000, 0.800000, 0.800000}
\pgfsetstrokecolor{dialinecolor}
\draw (28.700000\du,3.600000\du)--(30.450000\du,3.600000\du)--(30.450000\du,3.350000\du)--(32.200000\du,3.850000\du)--(30.450000\du,4.350000\du)--(30.450000\du,4.100000\du)--(28.700000\du,4.100000\du)--cycle;
\pgfsetbuttcap
\pgfsetmiterjoin
\pgfsetdash{}{0pt}
\definecolor{dialinecolor}{rgb}{0.800000, 0.800000, 0.800000}
\pgfsetstrokecolor{dialinecolor}
\draw (28.700000\du,3.600000\du)--(30.450000\du,3.600000\du)--(30.450000\du,3.350000\du)--(32.200000\du,3.850000\du)--(30.450000\du,4.350000\du)--(30.450000\du,4.100000\du)--(28.700000\du,4.100000\du)--cycle;
\definecolor{dialinecolor}{rgb}{0.000000, 0.000000, 0.000000}
\pgfsetstrokecolor{dialinecolor}
\node[anchor=east] at (42.725000\du,6.847500\du){$m \in M$};
\end{tikzpicture}}
	    }
		\end{center}
	\end{minipage}
	\begin{center}
       \subfloat[Data Transformation]{\label{fig:pm2apm_c}
          \scalebox{\tikzScale}{\adjustTikzSize \input{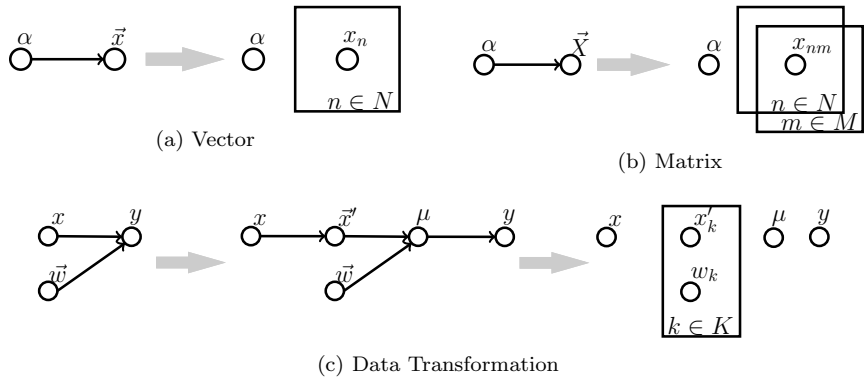}}
	    }
    \end{center}
	\caption[Conversion of Bayesian networks to atomic plate models]{
	Conversion of Bayesian networks to atomic plate models
	}
	\label{fig:pm2apm}
\end{figure}

Figures~\ref{fig:pm2apm_a} and \ref{fig:pm2apm_b} show the conversion of plate models containing a vector and a matrix variable, respectively.
The original multidimensional variable is represented in the APM by a plate containing an indexed component vertex.
The multidimensional variables, e.g. $\vec x\in\R^{|N|}$ and $\vec X\in\R^{|N|\times|M|}$, are converted into individual component variables $x_n$ and $x_{nm}$, surrounded by plates of the corresponding dimensions that are described by the respective index sets $N$ and $M$.

In this translation, edges are discarded.
If they would be preserved, they conveyed wrong semantics about the conditional probabilities of the BN after the decomposition.
However, it is not necessary to preserve this information in the conversion process to a database model, as it will not be used as a probabilistic model anymore.
By discarding all edges after the conversion process we avoid confusion as they might not represent conditional probability distributions in the APM.

Detecting the deterministic relationships and data transformations hidden in BNs is a bit more subtle.
Figure \ref{fig:pm2apm_c} illustrates this on the example of polynomial regression.
The one-dimensional input $x$ is transformed into a vector $\vec x'$ of several inputs $x_k'=x^k$ by taking different powers $k\in K\subset\Nat$.
All powers are multiplied with weights that are components of the vector $\vec w\in\R^{|K|}$.
Finally, the weighted powers are summed up and this sum is used as the mean $\mu$ of a normal distribution that governs a univariate random variable $y$.
As we introduced another multidimensional variable $\vec x'\in\R^{|K|}$, splitting all multidimensional variables yields a common plate with index set $K$ including all $x_k'$ and $w_k$.

\subsection{Translation of APM to ERM}
After converting a plate model to an APM, it is translated to an ERM.
For this step, we adapt the mapping from plate models to DAPER models \cite{heckerman2007probabilistic}.
We defer the discussion of the differences between our mapping and DAPER models to the related work section.
First, we state the basic translation rules.
Then, we consider some special cases of model constraints and show the influence on the resulting ERM.

\begin{figure}
		\centering
		\scalebox{\tikzScale}{\adjustTikzSize 
\ifx\du\undefined
  \newlength{\du}
\fi
\setlength{\du}{15\unitlength}
\begin{tikzpicture}
\pgftransformxscale{1.000000}
\pgftransformyscale{-1.000000}
\definecolor{dialinecolor}{rgb}{0.000000, 0.000000, 0.000000}
\pgfsetstrokecolor{dialinecolor}
\definecolor{dialinecolor}{rgb}{1.000000, 1.000000, 1.000000}
\pgfsetfillcolor{dialinecolor}
\pgfsetlinewidth{0.100000\du}
\pgfsetdash{}{0pt}
\pgfsetdash{}{0pt}
\pgfsetmiterjoin
\definecolor{dialinecolor}{rgb}{1.000000, 1.000000, 1.000000}
\pgfsetfillcolor{dialinecolor}
\fill (18.500000\du,4.500000\du)--(18.500000\du,10.500000\du)--(46.500000\du,10.500000\du)--(46.500000\du,4.500000\du)--cycle;
\definecolor{dialinecolor}{rgb}{1.000000, 1.000000, 1.000000}
\pgfsetstrokecolor{dialinecolor}
\draw (18.500000\du,4.500000\du)--(18.500000\du,10.500000\du)--(46.500000\du,10.500000\du)--(46.500000\du,4.500000\du)--cycle;
\pgfsetlinewidth{0.100000\du}
\pgfsetdash{}{0pt}
\pgfsetdash{}{0pt}
\pgfsetmiterjoin
\definecolor{dialinecolor}{rgb}{0.000000, 0.000000, 0.000000}
\pgfsetstrokecolor{dialinecolor}
\draw (19.000000\du,5.000000\du)--(19.000000\du,10.000000\du)--(26.500000\du,10.000000\du)--(26.500000\du,5.000000\du)--cycle;
\definecolor{dialinecolor}{rgb}{0.000000, 0.000000, 0.000000}
\pgfsetstrokecolor{dialinecolor}
\node[anchor=east] at (26.750000\du,9.416600\du){Objects ($n \in N$)};
\definecolor{dialinecolor}{rgb}{1.000000, 1.000000, 1.000000}
\pgfsetfillcolor{dialinecolor}
\pgfpathellipse{\pgfpoint{21.100000\du}{7.500000\du}}{\pgfpoint{0.500000\du}{0\du}}{\pgfpoint{0\du}{0.500000\du}}
\pgfusepath{fill}
\pgfsetlinewidth{0.100000\du}
\pgfsetdash{}{0pt}
\pgfsetdash{}{0pt}
\definecolor{dialinecolor}{rgb}{0.000000, 0.000000, 0.000000}
\pgfsetstrokecolor{dialinecolor}
\pgfpathellipse{\pgfpoint{21.100000\du}{7.500000\du}}{\pgfpoint{0.500000\du}{0\du}}{\pgfpoint{0\du}{0.500000\du}}
\pgfusepath{stroke}
\definecolor{dialinecolor}{rgb}{0.000000, 0.000000, 0.000000}
\pgfsetstrokecolor{dialinecolor}
\node[anchor=west] at (21.985400\du,7.630800\du){$x_n$};
\pgfsetlinewidth{0.100000\du}
\pgfsetdash{}{0pt}
\pgfsetdash{}{0pt}
\pgfsetbuttcap
\pgfsetmiterjoin
\pgfsetlinewidth{0.100000\du}
\pgfsetbuttcap
\pgfsetmiterjoin
\pgfsetdash{}{0pt}
\definecolor{dialinecolor}{rgb}{0.800000, 0.800000, 0.800000}
\pgfsetfillcolor{dialinecolor}
\fill (29.000000\du,7.250000\du)--(30.750000\du,7.250000\du)--(30.750000\du,7.000000\du)--(32.500000\du,7.500000\du)--(30.750000\du,8.000000\du)--(30.750000\du,7.750000\du)--(29.000000\du,7.750000\du)--cycle;
\definecolor{dialinecolor}{rgb}{0.800000, 0.800000, 0.800000}
\pgfsetstrokecolor{dialinecolor}
\draw (29.000000\du,7.250000\du)--(30.750000\du,7.250000\du)--(30.750000\du,7.000000\du)--(32.500000\du,7.500000\du)--(30.750000\du,8.000000\du)--(30.750000\du,7.750000\du)--(29.000000\du,7.750000\du)--cycle;
\pgfsetbuttcap
\pgfsetmiterjoin
\pgfsetdash{}{0pt}
\definecolor{dialinecolor}{rgb}{0.800000, 0.800000, 0.800000}
\pgfsetstrokecolor{dialinecolor}
\draw (29.000000\du,7.250000\du)--(30.750000\du,7.250000\du)--(30.750000\du,7.000000\du)--(32.500000\du,7.500000\du)--(30.750000\du,8.000000\du)--(30.750000\du,7.750000\du)--(29.000000\du,7.750000\du)--cycle;
\pgfsetlinewidth{0.100000\du}
\pgfsetdash{}{0pt}
\pgfsetdash{}{0pt}
\pgfsetmiterjoin
\definecolor{dialinecolor}{rgb}{1.000000, 1.000000, 1.000000}
\pgfsetfillcolor{dialinecolor}
\fill (35.000000\du,6.500000\du)--(35.000000\du,8.500000\du)--(40.000000\du,8.500000\du)--(40.000000\du,6.500000\du)--cycle;
\definecolor{dialinecolor}{rgb}{0.000000, 0.000000, 0.000000}
\pgfsetstrokecolor{dialinecolor}
\draw (35.000000\du,6.500000\du)--(35.000000\du,8.500000\du)--(40.000000\du,8.500000\du)--(40.000000\du,6.500000\du)--cycle;
\definecolor{dialinecolor}{rgb}{0.000000, 0.000000, 0.000000}
\pgfsetstrokecolor{dialinecolor}
\node at (37.500000\du,7.630800\du){Object};
\definecolor{dialinecolor}{rgb}{1.000000, 1.000000, 1.000000}
\pgfsetfillcolor{dialinecolor}
\pgfpathellipse{\pgfpoint{43.500000\du}{9.000000\du}}{\pgfpoint{2.500000\du}{0\du}}{\pgfpoint{0\du}{1.000000\du}}
\pgfusepath{fill}
\pgfsetlinewidth{0.100000\du}
\pgfsetdash{}{0pt}
\pgfsetdash{}{0pt}
\definecolor{dialinecolor}{rgb}{0.000000, 0.000000, 0.000000}
\pgfsetstrokecolor{dialinecolor}
\pgfpathellipse{\pgfpoint{43.500000\du}{9.000000\du}}{\pgfpoint{2.500000\du}{0\du}}{\pgfpoint{0\du}{1.000000\du}}
\pgfusepath{stroke}
\definecolor{dialinecolor}{rgb}{0.000000, 0.000000, 0.000000}
\pgfsetstrokecolor{dialinecolor}
\node at (43.500000\du,9.130800\du){$x$};
\pgfsetlinewidth{0.100000\du}
\pgfsetdash{}{0pt}
\pgfsetdash{}{0pt}
\pgfsetbuttcap
{
\definecolor{dialinecolor}{rgb}{0.000000, 0.000000, 0.000000}
\pgfsetfillcolor{dialinecolor}
\definecolor{dialinecolor}{rgb}{0.000000, 0.000000, 0.000000}
\pgfsetstrokecolor{dialinecolor}
\draw (40.049561\du,8.137390\du)--(41.331299\du,8.457825\du);
}
\definecolor{dialinecolor}{rgb}{1.000000, 1.000000, 1.000000}
\pgfsetfillcolor{dialinecolor}
\pgfpathellipse{\pgfpoint{43.500000\du}{6.000000\du}}{\pgfpoint{2.500000\du}{0\du}}{\pgfpoint{0\du}{1.000000\du}}
\pgfusepath{fill}
\pgfsetlinewidth{0.100000\du}
\pgfsetdash{}{0pt}
\pgfsetdash{}{0pt}
\definecolor{dialinecolor}{rgb}{0.000000, 0.000000, 0.000000}
\pgfsetstrokecolor{dialinecolor}
\pgfpathellipse{\pgfpoint{43.500000\du}{6.000000\du}}{\pgfpoint{2.500000\du}{0\du}}{\pgfpoint{0\du}{1.000000\du}}
\pgfusepath{stroke}
\definecolor{dialinecolor}{rgb}{0.000000, 0.000000, 0.000000}
\pgfsetstrokecolor{dialinecolor}
\node at (43.500000\du,6.130800\du){\underline{ID}};
\pgfsetlinewidth{0.100000\du}
\pgfsetdash{}{0pt}
\pgfsetdash{}{0pt}
\pgfsetbuttcap
{
\definecolor{dialinecolor}{rgb}{0.000000, 0.000000, 0.000000}
\pgfsetfillcolor{dialinecolor}
\definecolor{dialinecolor}{rgb}{0.000000, 0.000000, 0.000000}
\pgfsetstrokecolor{dialinecolor}
\draw (40.049561\du,6.862610\du)--(41.331299\du,6.542175\du);
}
\end{tikzpicture}}
		\caption[Translation of a plate containing a single variable to an ERM]{
			Translation of a plate containing a variable $x_n$ to an entity type having an artificial key ID and an attribute $x$.
		}
		\label{fig:pm2erm_uno_local}
	\end{figure}
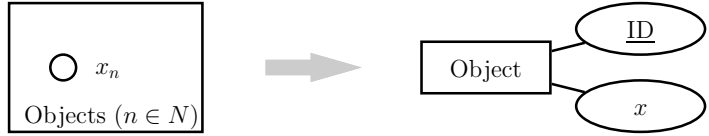

	\textbf{Translate plates to entity types.} Each plate is represented as an entity type.
	Usually there is an index set associated with each plate.
	Each entity type gets an artificial key (ID) that enumerates the index set.
	Figure~\ref{fig:pm2erm_uno_local} illustrates the translation of a plate, representing a set of Objects $N$, to the Object entity type.

\begin{figure}[t]
		\centering
		\scalebox{\tikzScale}{\adjustTikzSize \input{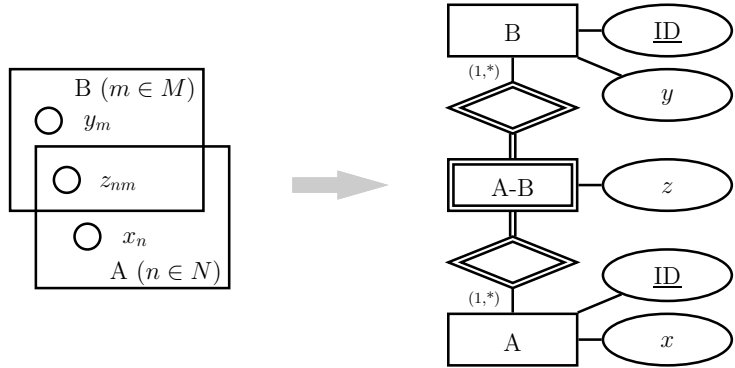}}
		\caption[Translation of a plate intersection to an ERM]{
			Translation of a plate intersection to an association entity of a many-to-many relationship.
			The variable $z_{nm}$ residing in the intersection is represented as the attribute $z$ of the association entity.
		}
		\label{fig:pm2erm_bi_noconstraints}
	\end{figure}

	\textbf{Translate plate intersections to relationships.} In general, plate intersections represent many-to-many relationships between the corresponding entity types.
	In contrast to \cite{heckerman2007probabilistic},  we express all relationships as association entity types
	\cite[pp.~86-88]{elmasri2007database}.
	This allows a generic translation procedure that can easily be used for n-ary relationships.
	Figure~\ref{fig:pm2erm_bi_noconstraints} shows the translation of two entity types having a binary relationship.

	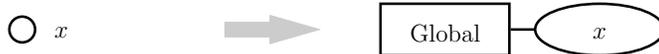
\begin{figure}[t]
		\centering
		\scalebox{\tikzScale}{\adjustTikzSize 
\ifx\du\undefined
  \newlength{\du}
\fi
\setlength{\du}{15\unitlength}
\begin{tikzpicture}
\pgftransformxscale{1.000000}
\pgftransformyscale{-1.000000}
\definecolor{dialinecolor}{rgb}{0.000000, 0.000000, 0.000000}
\pgfsetstrokecolor{dialinecolor}
\definecolor{dialinecolor}{rgb}{1.000000, 1.000000, 1.000000}
\pgfsetfillcolor{dialinecolor}
\pgfsetlinewidth{0.100000\du}
\pgfsetdash{}{0pt}
\pgfsetdash{}{0pt}
\pgfsetmiterjoin
\definecolor{dialinecolor}{rgb}{1.000000, 1.000000, 1.000000}
\pgfsetfillcolor{dialinecolor}
\fill (18.500000\du,4.500000\du)--(18.500000\du,10.500000\du)--(46.500000\du,10.500000\du)--(46.500000\du,4.500000\du)--cycle;
\definecolor{dialinecolor}{rgb}{1.000000, 1.000000, 1.000000}
\pgfsetstrokecolor{dialinecolor}
\draw (18.500000\du,4.500000\du)--(18.500000\du,10.500000\du)--(46.500000\du,10.500000\du)--(46.500000\du,4.500000\du)--cycle;
\definecolor{dialinecolor}{rgb}{1.000000, 1.000000, 1.000000}
\pgfsetfillcolor{dialinecolor}
\pgfpathellipse{\pgfpoint{21.100000\du}{7.500000\du}}{\pgfpoint{0.500000\du}{0\du}}{\pgfpoint{0\du}{0.500000\du}}
\pgfusepath{fill}
\pgfsetlinewidth{0.100000\du}
\pgfsetdash{}{0pt}
\pgfsetdash{}{0pt}
\definecolor{dialinecolor}{rgb}{0.000000, 0.000000, 0.000000}
\pgfsetstrokecolor{dialinecolor}
\pgfpathellipse{\pgfpoint{21.100000\du}{7.500000\du}}{\pgfpoint{0.500000\du}{0\du}}{\pgfpoint{0\du}{0.500000\du}}
\pgfusepath{stroke}
\definecolor{dialinecolor}{rgb}{0.000000, 0.000000, 0.000000}
\pgfsetstrokecolor{dialinecolor}
\node[anchor=west] at (21.985375\du,7.607875\du){$x$};
\pgfsetlinewidth{0.100000\du}
\pgfsetdash{}{0pt}
\pgfsetdash{}{0pt}
\pgfsetbuttcap
\pgfsetmiterjoin
\pgfsetlinewidth{0.100000\du}
\pgfsetbuttcap
\pgfsetmiterjoin
\pgfsetdash{}{0pt}
\definecolor{dialinecolor}{rgb}{0.800000, 0.800000, 0.800000}
\pgfsetfillcolor{dialinecolor}
\fill (29.000000\du,7.250000\du)--(30.750000\du,7.250000\du)--(30.750000\du,7.000000\du)--(32.500000\du,7.500000\du)--(30.750000\du,8.000000\du)--(30.750000\du,7.750000\du)--(29.000000\du,7.750000\du)--cycle;
\definecolor{dialinecolor}{rgb}{0.800000, 0.800000, 0.800000}
\pgfsetstrokecolor{dialinecolor}
\draw (29.000000\du,7.250000\du)--(30.750000\du,7.250000\du)--(30.750000\du,7.000000\du)--(32.500000\du,7.500000\du)--(30.750000\du,8.000000\du)--(30.750000\du,7.750000\du)--(29.000000\du,7.750000\du)--cycle;
\pgfsetbuttcap
\pgfsetmiterjoin
\pgfsetdash{}{0pt}
\definecolor{dialinecolor}{rgb}{0.800000, 0.800000, 0.800000}
\pgfsetstrokecolor{dialinecolor}
\draw (29.000000\du,7.250000\du)--(30.750000\du,7.250000\du)--(30.750000\du,7.000000\du)--(32.500000\du,7.500000\du)--(30.750000\du,8.000000\du)--(30.750000\du,7.750000\du)--(29.000000\du,7.750000\du)--cycle;
\pgfsetlinewidth{0.100000\du}
\pgfsetdash{}{0pt}
\pgfsetdash{}{0pt}
\pgfsetmiterjoin
\definecolor{dialinecolor}{rgb}{1.000000, 1.000000, 1.000000}
\pgfsetfillcolor{dialinecolor}
\fill (35.000000\du,6.500000\du)--(35.000000\du,8.500000\du)--(40.000000\du,8.500000\du)--(40.000000\du,6.500000\du)--cycle;
\definecolor{dialinecolor}{rgb}{0.000000, 0.000000, 0.000000}
\pgfsetstrokecolor{dialinecolor}
\draw (35.000000\du,6.500000\du)--(35.000000\du,8.500000\du)--(40.000000\du,8.500000\du)--(40.000000\du,6.500000\du)--cycle;
\definecolor{dialinecolor}{rgb}{0.000000, 0.000000, 0.000000}
\pgfsetstrokecolor{dialinecolor}
\node at (37.500000\du,7.630800\du){Global};
\definecolor{dialinecolor}{rgb}{1.000000, 1.000000, 1.000000}
\pgfsetfillcolor{dialinecolor}
\pgfpathellipse{\pgfpoint{43.500000\du}{7.500000\du}}{\pgfpoint{2.500000\du}{0\du}}{\pgfpoint{0\du}{1.000000\du}}
\pgfusepath{fill}
\pgfsetlinewidth{0.100000\du}
\pgfsetdash{}{0pt}
\pgfsetdash{}{0pt}
\definecolor{dialinecolor}{rgb}{0.000000, 0.000000, 0.000000}
\pgfsetstrokecolor{dialinecolor}
\pgfpathellipse{\pgfpoint{43.500000\du}{7.500000\du}}{\pgfpoint{2.500000\du}{0\du}}{\pgfpoint{0\du}{1.000000\du}}
\pgfusepath{stroke}
\definecolor{dialinecolor}{rgb}{0.000000, 0.000000, 0.000000}
\pgfsetstrokecolor{dialinecolor}
\node at (43.500000\du,7.630800\du){$x$};
\pgfsetlinewidth{0.100000\du}
\pgfsetdash{}{0pt}
\pgfsetdash{}{0pt}
\pgfsetbuttcap
{
\definecolor{dialinecolor}{rgb}{0.000000, 0.000000, 0.000000}
\pgfsetfillcolor{dialinecolor}
\definecolor{dialinecolor}{rgb}{0.000000, 0.000000, 0.000000}
\pgfsetstrokecolor{dialinecolor}
\draw (40.049561\du,7.500000\du)--(40.950439\du,7.500000\du);
}
\end{tikzpicture}}
		\caption[Translation of a variable associated with no plate to an ERM]{
			Translation of a single variable that is not located in any plate.
			It will be associated with the artificial entity type Global that has only one entity.
		}
		\label{fig:pm2erm_uno_global}
	\end{figure}

	\textbf{Translate variables to attributes.} The translation of attributes depends on the number of plates surrounding them.
	If a variable is surrounded by exactly one plate, the entity type of that plate gets an additional attribute representing this variable (see Figure~\ref{fig:pm2erm_uno_local}).
	If a variable resides inside multiple plates, it becomes an attribute of the corresponding association entity (see Figure~\ref{fig:pm2erm_bi_noconstraints}).
	Variables that are associated with no plate are assigned to an artificial entity type called Global (see Figure~\ref{fig:pm2erm_uno_global}).
	There exists only one entity of type Global.

	\begin{figure}[t]
		\centering
		\scalebox{\tikzScale}{\adjustTikzSize \input{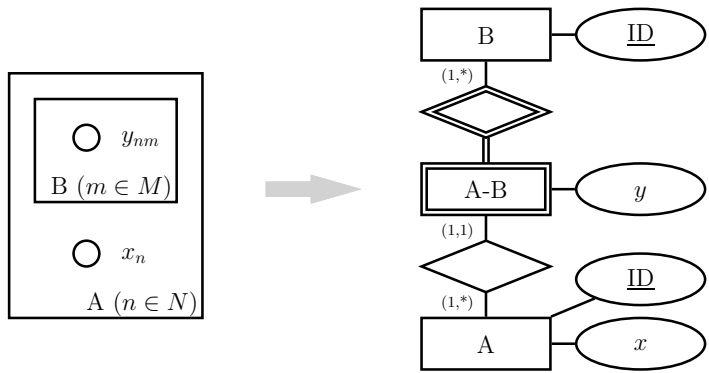}}
		\caption[Translation of nested plates to an ERM]{
			Translation of nested plates to an association entity.
			The entity type of the covering plate will have no weak relationship to the association entity.
			This reflects that every instance of B does belong to exactly one instance of A.
		}
		\label{fig:pm2erm_covered_1}
	\end{figure}

	\begin{figure}[t]
		\centering
		\scalebox{\tikzScale}{\adjustTikzSize \input{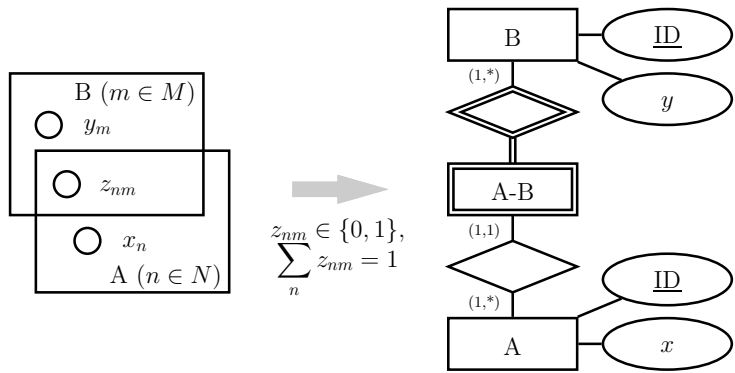}}
		\caption[Translation of plate intersections with 1-of-K variables to an ERM]{
			Translation of a plate intersection to an association entity given a 1-of-K relationship variable constraint ($z_{nm}$).
			The constraint ensures that every instance of B can only be connected to one instance of A.
			Instead of representing this variable directly, it will affect the cardinality of the resulting relationship.
			The entity type whose index set is summed over will not have a weak relationship.
		}
		\label{fig:pm2erm_bi_constraints}
	\end{figure}

  \textbf{Translate nested plates to one-to-many relationships.} If a plate is nested in another plate, the resulting relationship has one-to-many cardinality instead of many-to-many.
	This works well in the simple case of a binary relation, but falls short when one or both plates are additionally intersected or nested with further plates.
	Therefore, we adjust the mapping rule to the concept of association entities as demonstrated in Figure~\ref{fig:pm2erm_covered_1}, which works in the general case.
	In case of only binary relationships like in Figure~\ref{fig:pm2erm_covered_1}, the weak entity is not necessary.
	Such cases will be fixed in the reduction step that follows the translation to ERM.

Further, we propose two additional transformation rules that consider the effects of constraints for random variables on the resulting ERM and cope with self relationships as a result of matrix or tensor variable translations.

	\textbf{Adjust cardinalities depending on variable constraints.} In some applications there are variables that solely express associations between objects.
	Those variables can be coded by the 1-out-of-K scheme.
	This basically says that e.g. the vector $\vec z \in \{0,1\}^K$ contains only bits of zero and one and further obeys the additional summation constraint $\sum_{k=1}^K z_k = 1$, which says that exactly a single bit in the vector is set to one.
	In case such variable $z_k$ is in the intersection of two plates and there is such a summation constraint over either index set, then the plate intersection should not be translated as a many-to-many relationship but as one-to-many relationship.
    However, in this intermediate step of the translation, the one-to-many relationship is expressed by an association entity as in the original many-to-many relationship from the plate intersection to allow the translation to continue with further plate intersections covering $z_k$.
    In order to express the one-to-many constraint, the relationship between the entity whose index set is summed over and the association entity is translated as a normal relationship instead of a weak entity relationship.
    Finally, the constrained attribute $z_k$ of the association entity is removed since now the association information is expressed by the relationships as shown in Figure~\ref{fig:pm2erm_bi_constraints}.
	In case of only two overlapping plates the more complex expression of the one-to-many relationship is simplified in the final reduction step.

	\textbf{Translate overlapping plates with same index set to self relationships.} When converting plate models with matrix or tensor variables to APMs, the translation procedure will produce overlapping plates as illustrated in Figure~\ref{fig:pm2apm}b.
	Given a matrix with equal dimensions $N = M$, this will result in two overlapping plates of the same index set.
	These are simply translated as a many-to-many self relationship.
	The matrix components are then represented as an attribute of the resulting association entity.

With this set of rules it is possible to translate any given plate model into an ERM.
However, in some cases the resulting model might not be well-formed.
It might express simple one-to-many relationships in a complicated way.
Therefore, a reduction step is performed as described in the next section.

\subsection{Reduction of Translated Entity-Relationship Models}
\label{sec:erm_reduction}
In order to produce well-formed ERMs from plate models, it is necessary to apply a reduction step after the conversion from APM to ERM.
This reduction will make sure that all translation artifacts are eliminated that are caused by straight application of the translation rules.
Those artifacts are weak entities without primary key extension and duplicate relationships.

Weak entities without primary key extension may occur in cases like in Figure~\ref{fig:pm2erm_covered_1} and \ref{fig:pm2erm_bi_constraints}.
Plate intersections are translated to association entities having a weak relationship to all entities that form the intersection.
However, 1-of-K coded variables and nested plates turn weak entity relationships into normal ones.
In case only one weak relationship is left after the completed translation, the construct is not well-formed when it does not extend the primary key.
The weak entity relationship is then a degenerated one-to-one relationship and the weak entity is merged with parent entity.

\begin{figure}[t]
	\centering
	\scalebox{\tikzScale}{\adjustTikzSize \input{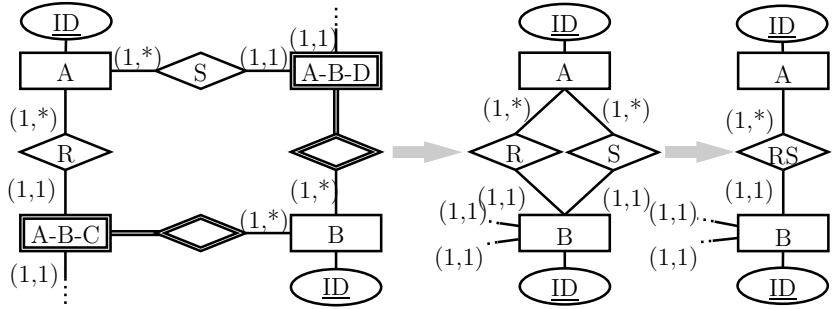}}
	\caption[Merging two reduced one-to-many relationships.]{
		Merging two reduced one-to-many relationships. After A-B-C and A-B-C are merged with B, there will be two relationships between A and B.
		We merge them if they are equivalent on a semantical level.
	}
	\label{fig:erm_reduction_minimal_example}
\end{figure}

After performing all possible merges as described above there may be duplicate relationships left.
Consider the case in Figure~\ref{fig:erm_reduction_minimal_example}.
Two weak entities are merged with A and B respectively.
Then, the relationships $R$ and $S$ may become redundant.
However, merging those is only permitted if the relationships are semantically expressing the same facts.
Thus, the translating person needs to evaluate the merging conditions depending on the concrete model.
There will be a practical example of this situation in the case study in Section~\ref{sec:casestudy}.

\section{Case Study: Text Topic Modeling}\label{sec:casestudy}
As a practical use case of our approach, we present TopicExplorer, an application to explore document collections using probabilistic topic models \cite{hinneburg2014topic,hinneburg2012topicexplorer}. 
In the first subsection, we explain LDA in detail, which is the probabilistic model behind TopicExplorer and demonstrate the translation of this plate model to an ERM. 
In the last part, we explain the use case and show typical analyses supported by the translated ERM.

\subsection{Translation of Latent Dirichlet Allocation Topic Model}
\begin{figure}[t]
\begin{minipage}[t]{0.49\linewidth}
	\begin{center}
		\subfloat[LDA plate model]{\label{fig:topic_platemodel_original}
    	\scalebox{\tikzScale}{\adjustTikzSize 
\ifx\du\undefined
  \newlength{\du}
\fi
\setlength{\du}{15\unitlength}
\begin{tikzpicture}
\pgftransformxscale{1.000000}
\pgftransformyscale{-1.000000}
\definecolor{dialinecolor}{rgb}{0.000000, 0.000000, 0.000000}
\pgfsetstrokecolor{dialinecolor}
\definecolor{dialinecolor}{rgb}{1.000000, 1.000000, 1.000000}
\pgfsetfillcolor{dialinecolor}
\pgfsetlinewidth{0.100000\du}
\pgfsetdash{}{0pt}
\pgfsetdash{}{0pt}
\pgfsetmiterjoin
\definecolor{dialinecolor}{rgb}{0.000000, 0.000000, 0.000000}
\pgfsetstrokecolor{dialinecolor}
\draw (22.850000\du,-1.500000\du)--(22.850000\du,1.950000\du)--(28.850000\du,1.950000\du)--(28.850000\du,-1.500000\du)--cycle;
\pgfsetlinewidth{0.100000\du}
\pgfsetdash{}{0pt}
\pgfsetdash{}{0pt}
\pgfsetmiterjoin
\definecolor{dialinecolor}{rgb}{0.000000, 0.000000, 0.000000}
\pgfsetstrokecolor{dialinecolor}
\draw (26.300000\du,2.350000\du)--(26.300000\du,7.700000\du)--(39.850000\du,7.700000\du)--(39.850000\du,2.350000\du)--cycle;
\pgfsetlinewidth{0.100000\du}
\pgfsetdash{}{0pt}
\pgfsetdash{}{0pt}
\pgfsetmiterjoin
\definecolor{dialinecolor}{rgb}{0.000000, 0.000000, 0.000000}
\pgfsetstrokecolor{dialinecolor}
\draw (30.800000\du,2.800000\du)--(30.800000\du,6.450000\du)--(38.900000\du,6.450000\du)--(38.900000\du,2.800000\du)--cycle;
\definecolor{dialinecolor}{rgb}{0.498039, 0.498039, 0.498039}
\pgfsetfillcolor{dialinecolor}
\pgfpathellipse{\pgfpoint{37.150000\du}{4.850000\du}}{\pgfpoint{0.500000\du}{0\du}}{\pgfpoint{0\du}{0.500000\du}}
\pgfusepath{fill}
\pgfsetlinewidth{0.100000\du}
\pgfsetdash{}{0pt}
\pgfsetdash{}{0pt}
\definecolor{dialinecolor}{rgb}{0.000000, 0.000000, 0.000000}
\pgfsetstrokecolor{dialinecolor}
\pgfpathellipse{\pgfpoint{37.150000\du}{4.850000\du}}{\pgfpoint{0.500000\du}{0\du}}{\pgfpoint{0\du}{0.500000\du}}
\pgfusepath{stroke}
\definecolor{dialinecolor}{rgb}{1.000000, 1.000000, 1.000000}
\pgfsetfillcolor{dialinecolor}
\pgfpathellipse{\pgfpoint{32.750000\du}{4.800000\du}}{\pgfpoint{0.500000\du}{0\du}}{\pgfpoint{0\du}{0.500000\du}}
\pgfusepath{fill}
\pgfsetlinewidth{0.100000\du}
\pgfsetdash{}{0pt}
\pgfsetdash{}{0pt}
\definecolor{dialinecolor}{rgb}{0.000000, 0.000000, 0.000000}
\pgfsetstrokecolor{dialinecolor}
\pgfpathellipse{\pgfpoint{32.750000\du}{4.800000\du}}{\pgfpoint{0.500000\du}{0\du}}{\pgfpoint{0\du}{0.500000\du}}
\pgfusepath{stroke}
\definecolor{dialinecolor}{rgb}{0.000000, 0.000000, 0.000000}
\pgfsetstrokecolor{dialinecolor}
\node at (37.150000\du,3.833700\du){$\vec d_{nm}$};
\definecolor{dialinecolor}{rgb}{0.000000, 0.000000, 0.000000}
\pgfsetstrokecolor{dialinecolor}
\node[anchor=east] at (38.900000\du,5.997500\du){Tokens ($M_n$)};
\definecolor{dialinecolor}{rgb}{0.000000, 0.000000, 0.000000}
\pgfsetstrokecolor{dialinecolor}
\node[anchor=east] at (39.850000\du,7.197500\du){Documents ($N$)};
\definecolor{dialinecolor}{rgb}{0.000000, 0.000000, 0.000000}
\pgfsetstrokecolor{dialinecolor}
\node at (32.750000\du,3.771250\du){$\vec z_{nm}$};
\definecolor{dialinecolor}{rgb}{1.000000, 1.000000, 1.000000}
\pgfsetfillcolor{dialinecolor}
\pgfpathellipse{\pgfpoint{28.100000\du}{4.850000\du}}{\pgfpoint{0.500000\du}{0\du}}{\pgfpoint{0\du}{0.500000\du}}
\pgfusepath{fill}
\pgfsetlinewidth{0.100000\du}
\pgfsetdash{}{0pt}
\pgfsetdash{}{0pt}
\definecolor{dialinecolor}{rgb}{0.000000, 0.000000, 0.000000}
\pgfsetstrokecolor{dialinecolor}
\pgfpathellipse{\pgfpoint{28.100000\du}{4.850000\du}}{\pgfpoint{0.500000\du}{0\du}}{\pgfpoint{0\du}{0.500000\du}}
\pgfusepath{stroke}
\definecolor{dialinecolor}{rgb}{0.000000, 0.000000, 0.000000}
\pgfsetfillcolor{dialinecolor}
\pgfpathellipse{\pgfpoint{23.950000\du}{4.800000\du}}{\pgfpoint{0.300000\du}{0\du}}{\pgfpoint{0\du}{0.300000\du}}
\pgfusepath{fill}
\pgfsetlinewidth{0.100000\du}
\pgfsetdash{}{0pt}
\pgfsetdash{}{0pt}
\definecolor{dialinecolor}{rgb}{0.000000, 0.000000, 0.000000}
\pgfsetstrokecolor{dialinecolor}
\pgfpathellipse{\pgfpoint{23.950000\du}{4.800000\du}}{\pgfpoint{0.300000\du}{0\du}}{\pgfpoint{0\du}{0.300000\du}}
\pgfusepath{stroke}
\definecolor{dialinecolor}{rgb}{0.000000, 0.000000, 0.000000}
\pgfsetstrokecolor{dialinecolor}
\node at (28.150000\du,3.841950\du){$\vec \theta_n$};
\definecolor{dialinecolor}{rgb}{0.000000, 0.000000, 0.000000}
\pgfsetstrokecolor{dialinecolor}
\node at (23.900000\du,3.841950\du){$\vec \alpha$};
\definecolor{dialinecolor}{rgb}{1.000000, 1.000000, 1.000000}
\pgfsetfillcolor{dialinecolor}
\pgfpathellipse{\pgfpoint{27.950000\du}{0.150000\du}}{\pgfpoint{0.500000\du}{0\du}}{\pgfpoint{0\du}{0.500000\du}}
\pgfusepath{fill}
\pgfsetlinewidth{0.100000\du}
\pgfsetdash{}{0pt}
\pgfsetdash{}{0pt}
\definecolor{dialinecolor}{rgb}{0.000000, 0.000000, 0.000000}
\pgfsetstrokecolor{dialinecolor}
\pgfpathellipse{\pgfpoint{27.950000\du}{0.150000\du}}{\pgfpoint{0.500000\du}{0\du}}{\pgfpoint{0\du}{0.500000\du}}
\pgfusepath{stroke}
\definecolor{dialinecolor}{rgb}{0.000000, 0.000000, 0.000000}
\pgfsetstrokecolor{dialinecolor}
\node at (28.000000\du,-0.658050\du){$\vec \mu_k$};
\definecolor{dialinecolor}{rgb}{0.000000, 0.000000, 0.000000}
\pgfsetstrokecolor{dialinecolor}
\node at (23.900000\du,-0.658050\du){$\vec \beta_k$};
\definecolor{dialinecolor}{rgb}{0.000000, 0.000000, 0.000000}
\pgfsetstrokecolor{dialinecolor}
\node[anchor=east] at (28.850000\du,1.447500\du){Topics ($K$)};
\definecolor{dialinecolor}{rgb}{0.000000, 0.000000, 0.000000}
\pgfsetfillcolor{dialinecolor}
\pgfpathellipse{\pgfpoint{23.850000\du}{0.150000\du}}{\pgfpoint{0.300000\du}{0\du}}{\pgfpoint{0\du}{0.300000\du}}
\pgfusepath{fill}
\pgfsetlinewidth{0.100000\du}
\pgfsetdash{}{0pt}
\pgfsetdash{}{0pt}
\definecolor{dialinecolor}{rgb}{0.000000, 0.000000, 0.000000}
\pgfsetstrokecolor{dialinecolor}
\pgfpathellipse{\pgfpoint{23.850000\du}{0.150000\du}}{\pgfpoint{0.300000\du}{0\du}}{\pgfpoint{0\du}{0.300000\du}}
\pgfusepath{stroke}
\pgfsetlinewidth{0.100000\du}
\pgfsetdash{}{0pt}
\pgfsetdash{}{0pt}
\pgfsetbuttcap
{
\definecolor{dialinecolor}{rgb}{0.000000, 0.000000, 0.000000}
\pgfsetfillcolor{dialinecolor}
\pgfsetarrowsend{to}
\definecolor{dialinecolor}{rgb}{0.000000, 0.000000, 0.000000}
\pgfsetstrokecolor{dialinecolor}
\draw (33.250000\du,4.800000\du)--(36.600610\du,4.842957\du);
}
\pgfsetlinewidth{0.100000\du}
\pgfsetdash{}{0pt}
\pgfsetdash{}{0pt}
\pgfsetbuttcap
{
\definecolor{dialinecolor}{rgb}{0.000000, 0.000000, 0.000000}
\pgfsetfillcolor{dialinecolor}
\pgfsetarrowsend{to}
\definecolor{dialinecolor}{rgb}{0.000000, 0.000000, 0.000000}
\pgfsetstrokecolor{dialinecolor}
\draw (28.650031\du,4.844086\du)--(32.199969\du,4.805914\du);
}
\pgfsetlinewidth{0.100000\du}
\pgfsetdash{}{0pt}
\pgfsetdash{}{0pt}
\pgfsetbuttcap
{
\definecolor{dialinecolor}{rgb}{0.000000, 0.000000, 0.000000}
\pgfsetfillcolor{dialinecolor}
\pgfsetarrowsend{to}
\definecolor{dialinecolor}{rgb}{0.000000, 0.000000, 0.000000}
\pgfsetstrokecolor{dialinecolor}
\draw (24.299548\du,4.804211\du)--(27.550348\du,4.843378\du);
}
\pgfsetlinewidth{0.100000\du}
\pgfsetdash{}{0pt}
\pgfsetdash{}{0pt}
\pgfsetbuttcap
{
\definecolor{dialinecolor}{rgb}{0.000000, 0.000000, 0.000000}
\pgfsetfillcolor{dialinecolor}
\pgfsetarrowsend{to}
\definecolor{dialinecolor}{rgb}{0.000000, 0.000000, 0.000000}
\pgfsetstrokecolor{dialinecolor}
\draw (24.199341\du,0.150000\du)--(27.400464\du,0.150000\du);
}
\pgfsetlinewidth{0.100000\du}
\pgfsetdash{}{0pt}
\pgfsetdash{}{0pt}
\pgfsetbuttcap
{
\definecolor{dialinecolor}{rgb}{0.000000, 0.000000, 0.000000}
\pgfsetfillcolor{dialinecolor}
\pgfsetarrowsend{to}
\definecolor{dialinecolor}{rgb}{0.000000, 0.000000, 0.000000}
\pgfsetstrokecolor{dialinecolor}
\draw (28.439929\du,0.400290\du)--(36.660071\du,4.599710\du);
}
\end{tikzpicture}}}
	\end{center}
\end{minipage}
\hspace{0.0cm}
\begin{minipage}[t]{0.49\linewidth}
	\begin{center}
		\subfloat[Atomic LDA plate model]{\label{fig:topic_platemodel_expanded}
		\scalebox{\tikzScale}{\adjustTikzSize 
\ifx\du\undefined
  \newlength{\du}
\fi
\setlength{\du}{15\unitlength}
\begin{tikzpicture}
\pgftransformxscale{1.000000}
\pgftransformyscale{-1.000000}
\definecolor{dialinecolor}{rgb}{0.000000, 0.000000, 0.000000}
\pgfsetstrokecolor{dialinecolor}
\definecolor{dialinecolor}{rgb}{1.000000, 1.000000, 1.000000}
\pgfsetfillcolor{dialinecolor}
\pgfsetlinewidth{0.100000\du}
\pgfsetdash{}{0pt}
\pgfsetdash{}{0pt}
\pgfsetmiterjoin
\pgfsetbuttcap
\definecolor{dialinecolor}{rgb}{0.000000, 0.000000, 0.000000}
\pgfsetstrokecolor{dialinecolor}
\draw (49.800000\du,23.650000\du)--(54.400000\du,23.600000\du)--(54.400000\du,12.900000\du)--(37.450000\du,12.850000\du)--(37.400000\du,16.850000\du)--(49.850000\du,16.800000\du)--cycle;
\definecolor{dialinecolor}{rgb}{0.000000, 0.000000, 0.000000}
\pgfsetstrokecolor{dialinecolor}
\node[anchor=east] at (54.700000\du,23.197500\du){Words ($V$)};
\pgfsetlinewidth{0.100000\du}
\pgfsetdash{}{0pt}
\pgfsetdash{}{0pt}
\pgfsetmiterjoin
\definecolor{dialinecolor}{rgb}{0.000000, 0.000000, 0.000000}
\pgfsetstrokecolor{dialinecolor}
\draw (37.070000\du,13.295000\du)--(37.070000\du,23.700000\du)--(49.137600\du,23.700000\du)--(49.137600\du,13.295000\du)--cycle;
\pgfsetlinewidth{0.100000\du}
\pgfsetdash{}{0pt}
\pgfsetdash{}{0pt}
\pgfsetmiterjoin
\definecolor{dialinecolor}{rgb}{0.000000, 0.000000, 0.000000}
\pgfsetstrokecolor{dialinecolor}
\draw (40.520000\du,17.145000\du)--(40.520000\du,22.495000\du)--(54.070000\du,22.495000\du)--(54.070000\du,17.145000\du)--cycle;
\pgfsetlinewidth{0.100000\du}
\pgfsetdash{}{0pt}
\pgfsetdash{}{0pt}
\pgfsetmiterjoin
\definecolor{dialinecolor}{rgb}{0.000000, 0.000000, 0.000000}
\pgfsetstrokecolor{dialinecolor}
\draw (43.537600\du,17.595000\du)--(43.537600\du,21.245000\du)--(53.120000\du,21.245000\du)--(53.120000\du,17.595000\du)--cycle;
\definecolor{dialinecolor}{rgb}{0.498039, 0.498039, 0.498039}
\pgfsetfillcolor{dialinecolor}
\pgfpathellipse{\pgfpoint{51.370000\du}{19.645000\du}}{\pgfpoint{0.500000\du}{0\du}}{\pgfpoint{0\du}{0.500000\du}}
\pgfusepath{fill}
\pgfsetlinewidth{0.100000\du}
\pgfsetdash{}{0pt}
\pgfsetdash{}{0pt}
\definecolor{dialinecolor}{rgb}{0.000000, 0.000000, 0.000000}
\pgfsetstrokecolor{dialinecolor}
\pgfpathellipse{\pgfpoint{51.370000\du}{19.645000\du}}{\pgfpoint{0.500000\du}{0\du}}{\pgfpoint{0\du}{0.500000\du}}
\pgfusepath{stroke}
\definecolor{dialinecolor}{rgb}{1.000000, 1.000000, 1.000000}
\pgfsetfillcolor{dialinecolor}
\pgfpathellipse{\pgfpoint{46.970000\du}{19.595000\du}}{\pgfpoint{0.500000\du}{0\du}}{\pgfpoint{0\du}{0.500000\du}}
\pgfusepath{fill}
\pgfsetlinewidth{0.100000\du}
\pgfsetdash{}{0pt}
\pgfsetdash{}{0pt}
\definecolor{dialinecolor}{rgb}{0.000000, 0.000000, 0.000000}
\pgfsetstrokecolor{dialinecolor}
\pgfpathellipse{\pgfpoint{46.970000\du}{19.595000\du}}{\pgfpoint{0.500000\du}{0\du}}{\pgfpoint{0\du}{0.500000\du}}
\pgfusepath{stroke}
\definecolor{dialinecolor}{rgb}{0.000000, 0.000000, 0.000000}
\pgfsetstrokecolor{dialinecolor}
\node at (51.370000\du,18.569406\du){$d_{nmv}$};
\definecolor{dialinecolor}{rgb}{0.000000, 0.000000, 0.000000}
\pgfsetstrokecolor{dialinecolor}
\node[anchor=east] at (49.470000\du,20.742500\du){Tokens ($M_n$)};
\definecolor{dialinecolor}{rgb}{0.000000, 0.000000, 0.000000}
\pgfsetstrokecolor{dialinecolor}
\node[anchor=east] at (47.620000\du,22.042500\du){Documents ($N$)};
\definecolor{dialinecolor}{rgb}{0.000000, 0.000000, 0.000000}
\pgfsetstrokecolor{dialinecolor}
\node at (46.970000\du,18.556906\du){$z_{nmk}$};
\definecolor{dialinecolor}{rgb}{1.000000, 1.000000, 1.000000}
\pgfsetfillcolor{dialinecolor}
\pgfpathellipse{\pgfpoint{42.320000\du}{19.645000\du}}{\pgfpoint{0.500000\du}{0\du}}{\pgfpoint{0\du}{0.500000\du}}
\pgfusepath{fill}
\pgfsetlinewidth{0.100000\du}
\pgfsetdash{}{0pt}
\pgfsetdash{}{0pt}
\definecolor{dialinecolor}{rgb}{0.000000, 0.000000, 0.000000}
\pgfsetstrokecolor{dialinecolor}
\pgfpathellipse{\pgfpoint{42.320000\du}{19.645000\du}}{\pgfpoint{0.500000\du}{0\du}}{\pgfpoint{0\du}{0.500000\du}}
\pgfusepath{stroke}
\definecolor{dialinecolor}{rgb}{0.000000, 0.000000, 0.000000}
\pgfsetfillcolor{dialinecolor}
\pgfpathellipse{\pgfpoint{38.170000\du}{19.595000\du}}{\pgfpoint{0.300000\du}{0\du}}{\pgfpoint{0\du}{0.300000\du}}
\pgfusepath{fill}
\pgfsetlinewidth{0.100000\du}
\pgfsetdash{}{0pt}
\pgfsetdash{}{0pt}
\definecolor{dialinecolor}{rgb}{0.000000, 0.000000, 0.000000}
\pgfsetstrokecolor{dialinecolor}
\pgfpathellipse{\pgfpoint{38.170000\du}{19.595000\du}}{\pgfpoint{0.300000\du}{0\du}}{\pgfpoint{0\du}{0.300000\du}}
\pgfusepath{stroke}
\definecolor{dialinecolor}{rgb}{0.000000, 0.000000, 0.000000}
\pgfsetstrokecolor{dialinecolor}
\node at (42.370000\du,18.577606\du){$\theta_{nk}$};
\definecolor{dialinecolor}{rgb}{0.000000, 0.000000, 0.000000}
\pgfsetstrokecolor{dialinecolor}
\node at (38.120000\du,18.627606\du){$\alpha_k$};
\definecolor{dialinecolor}{rgb}{1.000000, 1.000000, 1.000000}
\pgfsetfillcolor{dialinecolor}
\pgfpathellipse{\pgfpoint{42.170000\du}{14.945000\du}}{\pgfpoint{0.500000\du}{0\du}}{\pgfpoint{0\du}{0.500000\du}}
\pgfusepath{fill}
\pgfsetlinewidth{0.100000\du}
\pgfsetdash{}{0pt}
\pgfsetdash{}{0pt}
\definecolor{dialinecolor}{rgb}{0.000000, 0.000000, 0.000000}
\pgfsetstrokecolor{dialinecolor}
\pgfpathellipse{\pgfpoint{42.170000\du}{14.945000\du}}{\pgfpoint{0.500000\du}{0\du}}{\pgfpoint{0\du}{0.500000\du}}
\pgfusepath{stroke}
\definecolor{dialinecolor}{rgb}{0.000000, 0.000000, 0.000000}
\pgfsetstrokecolor{dialinecolor}
\node at (42.220000\du,14.077606\du){$\mu_{kv}$};
\definecolor{dialinecolor}{rgb}{0.000000, 0.000000, 0.000000}
\pgfsetstrokecolor{dialinecolor}
\node at (38.320000\du,14.127606\du){$\beta_{kv}$};
\definecolor{dialinecolor}{rgb}{0.000000, 0.000000, 0.000000}
\pgfsetstrokecolor{dialinecolor}
\node[anchor=east] at (49.137600\du,23.197500\du){Topics ($K$)};
\definecolor{dialinecolor}{rgb}{0.000000, 0.000000, 0.000000}
\pgfsetfillcolor{dialinecolor}
\pgfpathellipse{\pgfpoint{38.070000\du}{14.945000\du}}{\pgfpoint{0.300000\du}{0\du}}{\pgfpoint{0\du}{0.300000\du}}
\pgfusepath{fill}
\pgfsetlinewidth{0.100000\du}
\pgfsetdash{}{0pt}
\pgfsetdash{}{0pt}
\definecolor{dialinecolor}{rgb}{0.000000, 0.000000, 0.000000}
\pgfsetstrokecolor{dialinecolor}
\pgfpathellipse{\pgfpoint{38.070000\du}{14.945000\du}}{\pgfpoint{0.300000\du}{0\du}}{\pgfpoint{0\du}{0.300000\du}}
\pgfusepath{stroke}
\end{tikzpicture}}}
	\end{center}
\end{minipage}\\
\caption[Transformation of LDA plate model to an APM]{Transformation of the LDA plate model to an APM.
}
\label{img:topic_platemodels}
\end{figure}
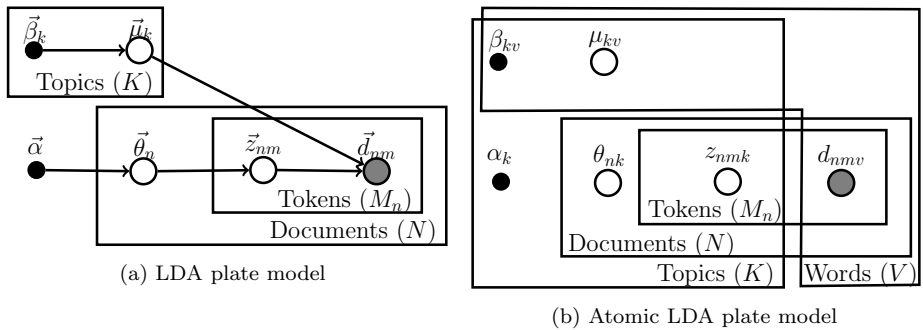

LDA models a collection of documents that is indexed by the set $N$.
Each document $n \in N$ consists of a set of tokens $M_n$ that represents the words occurring in this document.
The document specific token index sets $M_n$ partition the total index set of tokens $M$.
A token $m \in M_n$ correspondes to exactly one word type $v$ from a vocabulary $V$. 
In the Bayesian network, Figure~\ref{fig:topic_platemodel_original}, this information is coded as a bit vector $\vec d_{nm}\in\{0,1\}^{|V|}$ that has exactly a single 1 at the index associated with the respective word $v\in V$. 
Each token $m \in M_n$ is also assigned to a topic $k \in K$.
This assignment is coded by the bit vector $\vec z_{nm}\in\{0,1\}^{|K|}$, which has a single 1 at the respective topic index. 
The constraints of the single 1 in both bit vectors are ensured by requiring that the sums of the respective vector elements are one.
Furthermore, each topic has its own word distribution parameterized by a vector of positive real number $\vec \mu_k\in\R^{|V|}$. 
The topic proportions per document are represented by a similar vector $\vec \theta_n\in\R^{|K|}$.
Both vector also sum to one, however, in contrast to the bit vectors, these constraints have no special impact during the translation of the Bayesian network as they are vectors of real numbers.
The hyper-parameter vector $\vec \alpha\in\R^{|K|}$ and $\vec \beta_k\in\R^{|V|}$ regulate the prior distributions for the respecitve hidden parameters $\vec \theta_n$ and $\vec \mu_k$.

To obtain an ERM for LDA we follow the steps described in Section~\ref{sec:pm2erm}. 
First, the LDA plate model is transformed into an APM, secondly, this APM is translated to an ERM and last, the ERM is reduced to the final outcome.
The transformation to an APM is illustrated in Figure~\ref{img:topic_platemodels}. 
All vector variables are explicitly modeled as repeated variables for their components. 
The \verb|topic| plate now covers the components of the $|K|$-dimensional vector variables $\vec z_{nm}, \vec \theta_n$ and $\vec \alpha$. 
A new plate is created for the dimensionality $|V|$ to model the components of $\vec d_{nm}, \vec \mu_k$ and $\vec \beta_k$, representing the vocabulary of the documents.

\begin{figure}[t]
\centering
\scalebox{0.5}{\adjustTikzSize \input{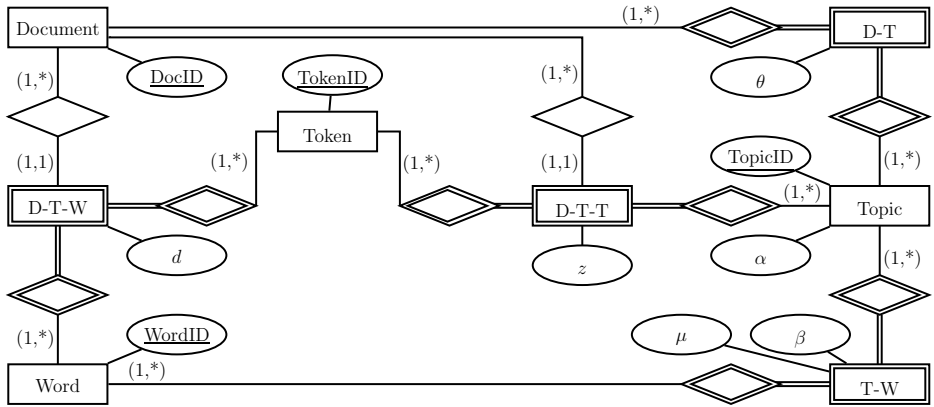}}
\caption[Intermediate ERM for LDA, translated from an APM]{First intermediate ERM for LDA translated from an APM. The entities topic and word still yields non-weak relationships to D-T-T and D-T-W respectively.
}\label{fig:topic_erm_good_verbose}
\end{figure}

\begin{figure}[t]
\centering
\scalebox{0.5}{\adjustTikzSize \input{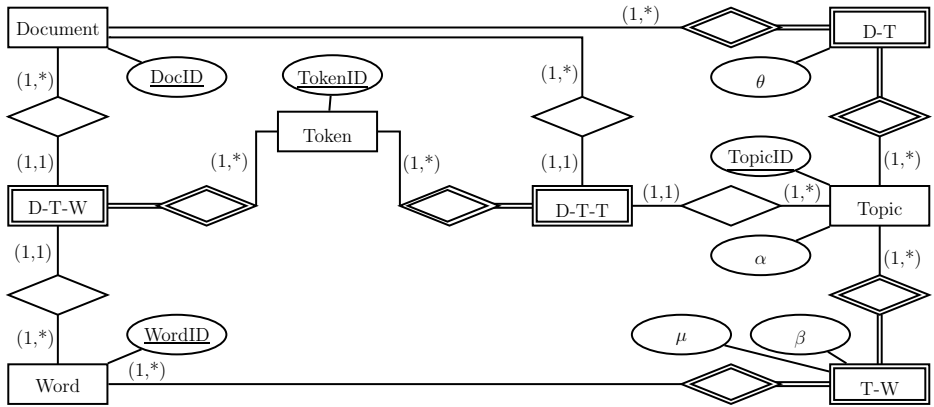}}
\caption[LDA ERM after taking constraints into account]{Second intermediate ERM for LDA after adjusting cardinalities and relationships according to the given variable constraints.}\label{fig:topic_erm_good_constraints}
\end{figure}

In the next step, the APM is translated to an intermediate ERM. 
Figure~\ref{fig:topic_erm_good_verbose} shows the result. 
The entities \verb|Document|, \verb|Word|, \verb|Token|, and \verb|Topic| correspond to the respective plates. 
The association entities \verb|D-T-W|, \verb|D-T-T|, \verb|D-T|, and \verb|T-W| represent the plate intersections. 
The \verb|Token| plate is nested inside the \verb|Document| plate, yielding non-weak relationships from \verb|Document| to \verb|D-T-W| and \verb|D-T-T|.
Furthermore, the summation constraints on the bit vector variables $\vec d_{nm}$ and $\vec z_{nm}$ are considered in the intermediate ERM in Figure~\ref{fig:topic_erm_good_constraints}. 
This yields a transformation of the relationships between \verb|Word| and \verb|D-T-W|, and \verb|Topic| and \verb|D-T-T|, respectively. 
These relationships are no longer weak and do not contribute to the primary key of the corresponding association entity.
Finally, the intermediate ERM is reduced to the result as shown in the introduction, Figure~\ref{fig:ex1_er}(right). 
In particular, the \verb|Token| entity type is merged with \verb|D-T-W| and \verb|D-T-T|. 
The relationships linking to these to entities are then connected to \verb|Token|.
Thus, there are now two relationship between Document and Token.
As these relationships have the same semantics, they are merged into a single one. 

The reduced, final ERM is exactly as one would expect it when designing a data model for LDA from scratch.
A document consists of one or more tokens which are of exactly one word type. 
Each token is assigned to a topic, while one topic can have multiple tokens assigned. 
The inferred topic mixture for each document is stored in \verb|D-T.|$\theta$, while \verb|T-W.|$\mu$ holds the word probabilities for each topic. 
The hyper-parameter $\alpha$ of the prior for the topic mixture resides as an attribute of the Topic entity type. 
The parameters for the individual priors for the word distributions are stored in \verb|T-W.|$\beta$.

\subsection{TopicExplorer}
TopicExplorer is a web application that helps users from the humanities to work with topic models, e.g. in a collaboration with the institute for Japanese studies at Martin-Luther-University, we analyzed blog posts about the Fukushima disaster. 
After crawling relevant blogs, each blog entry is preprocessed by computer linguistic software,   which is MeCab (\texttt{https://sourceforge.net/projects/mecab/}) in this case, to extracts tokens from full text.
The tokens are stored in their lemmatized form together with their part-of-speech tags (e.g. noun, verb or adjective) and their positions in the text. 
The full text of each document is stored in raw format as well to be able to view it in the user interface. 
Additionally, the blog post title, URL, and publication date are stored. 
\label{subsec:casestudy_analysis}

TopicExplorer allows user to analyze the topic structure of the documents using interactive visualizations.
The visualizations require analysis steps on the given data about the documents join together with results from the topic model.
The integrated ERM, see Figure \ref{fig:ex1_er}, gave the application developer a good idea how to access those data, without needing to understand the machine learning details of a topic model.
We present how to derive a few visulizations that are part of the current version of TopicExplorer \cite{hinneburg2014topic}.

\noindent\textbf{Document topic mixture.} LDA assigns a vector of topic probabilities stored in \verb|D-T.|$\theta$ for each document, called a topic mixture. This could either be presented as a list of topics with decreasing order of probabilities or as a pie chart like in the Topic Model Visualization Engine \cite{chaney2012visualizing}.

\noindent\textbf{Topic Documents.} Reversing the idea behind the document topic mixture, one can visualize a topic as a list of the most representative documents for this topic. This is done by joining \verb|Document|, \verb|Token| and \verb|Topic|, grouping by both IDs of documents and topics and counting the number of tokens in each group. For each topic the entries are sorted with decreasing token count, yielding a list of representative documents.

\noindent\textbf{Topic words.} As stated above, a topic can be represented as a list of words sorted in descreasing probability (\verb|T-W.|$\mu$). This list can be cut off, yielding a top words visualization for each topic.

\noindent\textbf{Topic frames.} Another visualization of topics uses the concept of frames. 
A \emph{topic frame} consists of a noun and a verb that are assigned to the same topic and appear close together in the same documents.
Thus, instead of visualizing a topic as top words, top frames can be used. 
Topic frames can be computed using \verb|Token|, \verb|Word| and \verb|Topic|, grouping by topic ID and word IDs of the frame tokens, and counting the number of frames having equal word types per topic. 
\noindent\textbf{Topic time.} To analyze how the discussion about Fukushima develops, one can visualize how the topics change over time. It is possible to analyze the dynamics of the topics in a postprocessing step after using LDA. To achieve this, entries are grouped depending on their publication date, e.g. monthly. Given those two attributes \verb|Document.Year| and \verb|Document.Month|\footnote{Depending on the implementation of the database, those attributes can be extracted on the fly out of the stored timestamp.}, the \verb|Document| table joined with \verb|Topic| is grouped by the \verb|TopicID|, \verb|Document.Year| and \verb|Document.Month|. Finally, counting the tokens assigned to each topic at each point of time allows plotting a time line for each topic.

\section{Conceptual Modeling Framework}\label{sec:framework}
\begin{figure}[t]
  \scalebox{0.45}{\adjustTikzSize \input{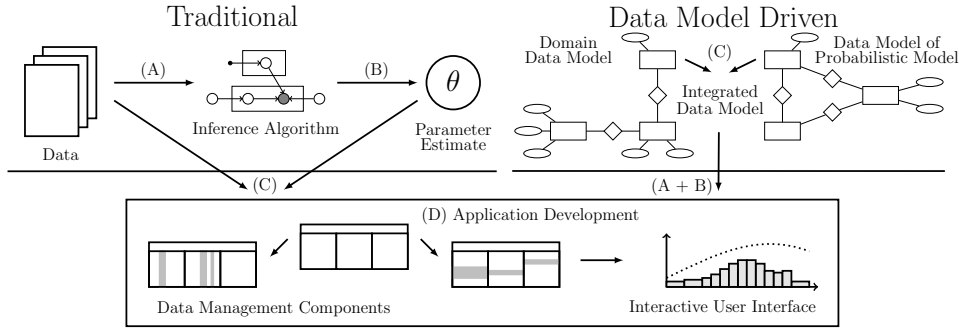}}
\caption[Traditional development versus data model driven development]{Comparison of traditional development versus data model driven development of big data analysis applications.}
\label{img:architecture}
\end{figure}
Based on the new method for translating probabilistic models to ERMs, we propose a first idea for a new data model driven development approach dedicated to big data analytics applications.
Figure~\ref{img:architecture} visualizes the traditional and our proposed data model driven development approaches.
Both address four different tasks, namely (A) gather the data sources and make them available to a probabilistic model, (B) run machine learning components, (C) integrate the data sources with the machine learning output and (D) build the application consisting of data management components and an interactive user interface.

The traditional approach addresses the tasks mainly in sequential order.
The first three steps implement data mining process following the CRISP model\cite{Han:2011:DMC:1972541}, while the last step addresses standard application development.

The translation method from BN to ERM allows an alternative, data model driven approach.
We assume that for a wide spectrum of machine learning problems abstract, readily developed BNs already do exist.
Those BNs could be pre-translated to ERMs to build a library.
Thus, conceptual information about the machine learning component is already available when integrating the data sources, task (C).
The BN could be treated as just another data source.
The entities corresponding to observed variables in the BN, including their respective relationships, have to be matched with those from other available data sources.
The matching conceptually defines the interface between data sources and machine learning, task (A).
Furthermore, translating the integrated ERM into a (relational) model for a big data framework, e.g. Spark \cite{Armbrust:2015:SSR:2723372.2742797}, conceptually defines the API between the output of machine learning and the rest of the application, task (B).
Depending on the framework, the tasks (A) and (B) could be supported by generation of efficient code for interfaces between data management and machine learning implementations.

As a consequence, the application developers just need knowledge about in- and outputs, and the relationships among the variables in the Bayesian model, but not about probabilistic distributions and dependencies.
Thus, our new data model driven approach eliminates unnecessary complexity caused by a lack of compatible conceptual languages on both sides of machine learning and data management.
Thereby, it makes the collaboration between machine learning experts and the application developers more direct and offers potential for efficiency optimization.

\section{Related Work}\label{sec:relwork}
Directed acyclic probabilistic entity-relationship (DAPER) models \cite{heckerman2007probabilistic} are closely related to our work.
However, they are designed to unify probabilistic relational models and plate models.
While we discard all facts about probablistic relationships in our translation, DAPER models can still be used to express such relationships.
As a consequence, vector variables and other implicit information is not resolved in DAPER models like in the proposed translation to APMs.
Thus, DAPER models are not used and also not intended to work as ERMs for conceptional design of data management.

There are several approaches that combine data management with machine learning, however, none of them reaches a comparable conceptual level like ERMs.
Hazy \cite{kumar2013hazy} provides programming, infrastructure and statistical processing abstractions, the latter are based on Markov logic \cite{domingos2007markov}.
This requires a deeper understanding of the machine learning algorithms in order to combine them effectively with data management.
Several approaches combine machine learning APIs with SQL \cite{akdere2011case,hellerstein2012madlib,Armbrust:2015:SSR:2723372.2742797} or with their own declarative language \cite{sparks2013mli}.

Last, data management is combined with machine learning at the level of user interfaces.
Examples are  Weka \cite{hall2009weka} and scikit-learn \cite{scikit-learn}, which  enable users to quickly select data sources and try different algorithms.
Both do not offer an easy way to integrate machine learning results with domain specific meta data.
Our approach also contrasts with statistical programming languages and software like R and SAS that just offer programming APIs to data sources and machine learning algorithms.

Our work is closely related in spirit to a recently proposed conceptual modeling framework work for network analytics \cite{Wang201559}.
In contrast, we target BNs, but we believe that our vision of a conceptual framework for probabilistic models can yield similar benefits.

\section{Conclusion}\label{sec:conclusion}
Our proposed translation shows that modeling an ERM for a given BN is a non-trival task.
Knowledge of the translation procedure helps data architects to pose the right questions for machine learning experts to uncover implicit information in a BN.
The subsequently proposed framework gives guidelines how to effectively build an integrated conceptual model that includes details about domain specific aspects as well as the machine learning side of a big data analytics application.
Future work includes the implementation of the framework and optimizing efficiency when translating integrated conceptual models to a particular implementation.

\bibliographystyle{plain}
\bibliography{literature}

\end{document}